\documentclass[useAMS,usenatbib]{mn2e}

\usepackage{graphicx}
\usepackage{epsfig}
\usepackage{amsmath}
\usepackage{amssymb}
\usepackage{subfigure}
\usepackage{epsfig}
\usepackage{array}
\usepackage{lscape}

\def\nat{\reff@jnl{Nature}}  
\def\reff@jnl#1{{\rm#1\/}}
\def\aj{\reff@jnl{AJ}}                  
\def\araa{\reff@jnl{ARA\&A}}            
\def\apj{\reff@jnl{ApJ}}                
\def\apjl{\reff@jnl{ApJ}}               
\def\apjs{\reff@jnl{ApJS}}     
\def\aap{\reff@jnl{A\&A}}  
\def\mnras{\reff@jnl{MNRAS}}  
\def\pasj{\reff@jnl{PASJ}}              

\begin{document}

\newcommand{\cl}{Cl\,0016+16}
\newcommand{\four}{MS\,0451.6-0305}
\newcommand{\ten}{MS\,1054.4-0321}
\newcommand{\rx}{RXJ\,2129.6+0005}

\title[]{Preliminary Sunyaev Zel'dovich Observations of Galaxy
Clusters with OCRA-p}

\author[]{Katy Lancaster$^1$, Mark Birkinshaw$^1$, Marcin
  P. Gawro\'nski$^3$, Ian Browne$^2$,
\newauthor Roman Feiler$^3$,  Andrzej Kus$^3$, 
Stuart Lowe$^2$, Eugeniusz Pazderski$^3$
\newauthor and Peter Wilkinson$^2$\\ 
$^1$ University of Bristol, Tyndall Avenue, Bristol
BS6 5BX\\ 
$^2$ Jodrell Bank Observatory, University of Manchester,
Macclesfield, Cheshire SK11 9DL\\ 
$^3$ Torun Centre for Astronomy,
Nicolas Copernicus University, ul. Gagarina 11, 87-100 Torun, POLAND\\
}

\date{Received **insert**; Accepted **insert**}

\pagerange{\pageref{firstpage}--\pageref{lastpage}} 
\pubyear{}

\maketitle
\label{firstpage}

\begin{abstract}
We present 30\,GHz Sunyaev Zel'dovich (SZ) observations of a sample of
four galaxy clusters with a prototype of the One Centimetre Receiver
Array (OCRA-p) which is mounted on the Torun 32-m telescope.  The
clusters (\cl, \four, \ten\space and Abell 2218) are popular SZ
targets and serve as commissioning observations.  All four are
detected with clear significance ($4-6\sigma$) and values for the
central temperature decrement are in good agreement with measurements
reported in the literature.  We believe that systematic effects are
successfully suppressed by our observing strategy.  The relatively
short integration times required to obtain these results demonstrate
the power of OCRA-p and its successors for future SZ studies.
\end{abstract}



\section{INTRODUCTION}
\label{intro}

The SZ effect is a spectral distortion of the Cosmic Microwave
Background (CMB) caused by the inverse Compton scattering of CMB
photons off the hot plasma found in clusters of galaxies.  At low
radio frequencies the SZ effect manifests itself as a fractional
decrement in the CMB of order $10^{-4}$, whereas at frequencies
greater than $\sim 220$\,GHz an increment is observed.  

Since the first SZ detections in the 1970s (\citealp{Birkinshaw1978a})
much progress has been made.  Extensive interferometric studies have
been undertaken, from the first SZ images (\citealp{Jones1993}), to
the use of cluster samples to perform cosmological studies
(e.g. \citealp{Grego2001}, \citealp{Reese2002},
\citealp{Saunders2003}, \citealp{Jones2005} and
\citealp{Bonamente2006}).  Radiometers (e.g. \citealp{Myers_obs}) and
bolometers (e.g. \citealp{Benson2004}) have also been used to good
effect.  More recently, purpose--built CMB instruments have proved
their worth in SZ studies (\citealp{Lancaster2005},
\citealp{Udomprasert2004}), although contamination from primordial
anisotropies can severely limit the achievable signal-to-noise ratio.

Measurements of the thermal and kinematic SZ effects are studied for
the unique information that they can provide on cosmology and the
structures of cluster atmospheres (see reviews of
\citealp{Birkinshaw1999} and \citealp{Carlstrom2002}).  Although most
work in the field of SZ studies to date has been performed using
non--ideal instruments, numerous purpose--built SZ observatories are
currently under construction (e.g. AMiBA \citep{Lo2000}, AMI
\citep{Kneissl2001}, SZA \citep{Loh2005}, ACT \citep{Kosowsky2003} and
SPT \citep{Ruhl2004}) all with the common aim of performing blind surveys.
Such work will exploit the unique redshift--independence of the SZ
effect to produce catalogues which are essentially mass-limited, and
thus less affected by the biases which plague selection via optical or
X-ray methods.  Studies of number counts will be used to further
constrain the cosmological model and study cluster
evolution from formation until the present.  Those telescopes which
have the capability to produce high--quality imaging will additionally
provide the opportunity to better understand cluster physics.

OCRA (\citealp{Browne2000}) is a planned 100-element continuum
receiver which will have excellent surveying and imaging
capabilities, and will thus be ideal for performing blind surveys in
order to study radio source populations and SZ clusters.  The
prototype for OCRA, OCRA-p, is a two--element receiver mounted on the
32-m telescope at the Torun Centre for Astrophysics of the Nicolas
Copernicus University, Poland.  Even at this preliminary stage, while
the instrument concept is being tested, high--sensitivity SZ
measurements are possible, with imaging and surveying planned for the
future.  We here present our first observations of 4 well-studied
massive SZ clusters (\cl, \four, \ten\space and A2218).

The structure of this paper is as follows.  In Section 2 we give a
brief description of the Torun 32-m telescope and the OCRA-p receiver.
Section 3 contains details of the cluster sample.  The observations
and data reduction are described in Section 4, and the problem of
radio source contamination is detailed in Section 5.  Section 6
presents our results and discussion.  Throughout the paper
we adopt the following cosmological parameter values: $H_0 = 70
\, \mathrm{km} \, \mathrm{s}^{-1} \, \mathrm{Mpc}^{-1}$,
$\Omega_{\mathrm{m}0} = 0.3$, $\Omega_{\Lambda 0} = 0.7$


\section{THE TORUN TELESCOPE AND OCRA-p}

The Torun observatory is located in Piwnice, 15\,km outside Torun in
northern Poland.  The telescope consists of a 32-m parabolic dish and
3.2-m sub-reflector, with a fully steerable classical Alt--Az mount.
It has receivers operating at 1.4-1.7, 5, 6.8, 12 and now, with
OCRA-p, 30\,GHz.  The telescope is used for a variety of studies
including interstellar molecules (e.g. \citealp{Blaszkiewicz2004}),
and VLBI (e.g. \citealp{Bartkiewicz2005}).

The OCRA prototype, OCRA-p, is funded by a grant from the Royal
Society Paul Instrument fund.  The instrument is described in detail
in \cite{LowePhD} and \cite{Lowe2007}; here we present a short
summary.  The basic radiometer design is based on the prototype for
the Planck Low Frequency Instrument (LFI, \cite{Mandolesi2000}), and
is similar to the WMAP K-band receivers \citep{Jarosik2003}.  OCRA-p
consists of two horn--feeds, the beams of which are separated by
$3'.1$ and have FWHM $1'.2$.  As the beam separation is small, it is
possible to reduce the effects of atmospheric and gain fluctuations by
switching between the input channels and taking the difference.  This
can be improved upon by further levels of switching.  The full
switching strategy for SZ observations is described in Section
\ref{sec:obs}.


\section{THE CLUSTER SAMPLE}
\label{sec:clusters}

For these preliminary observations, we have chosen four clusters which
have well--studied SZ effects, and which cover a wide range of
declinations, in order to assess the performance of the
telescope/receiver combination.  In a future paper we will present
results from observations of a large, statistically--complete sample
of clusters, observations of which are underway. The cluster sample,
and its X-ray properties, is summarised in Table \ref{tab:coord}.

\begin{table*}
\begin{tabular}{lccccr@{}lc}
\hline
Cluster &RA &DEC &$z$ &$L_\mathrm{x}$ &\multicolumn{2}{c}{$T_\mathrm{x}$}
 &References\\
 &(J2000) &(J2000) & &($10^{44}$erg\,s$^{-1}$) &\multicolumn{2}{c}{(keV)}\\
\hline
\cl &00 18 33.30   &+16 26 36.00 &0.546 &22.7 &$9.13${}&$^{+0.24}_{-0.22}$
&j, h, i \\
\four &04 54 10.90 &-03 01 07.00 &0.550 &17.5 &$8.05${}&$^{+0.5}_{-0.4}$
&b, f, g \\
\ten &10 57 00.20  &-03 37 27.00 &0.823 &12.3
&$7.2${}&$^{+0.7}_{-0.6}$ &b, e, e\\
A2218 &16 35 49.10 &+66 12 30.00 &0.176 &9.6
&$6.63${}&$^{+0.27}_{-0.27}$ &a, c, d\\
\hline
\end{tabular}
\caption{Basic cluster data. Coordinates correspond to pointing
  centres, parameters are as listed in the BAX X-ray cluster database.
  Luminosities refer to the 0.1-2.4\,keV ROSAT energy band.
  References: (a) \protect\citealp{Struble1999}, (b)
  \protect\citealp{Gioia1994}, (c) \protect\citealp{Balogh2002}, (d)
  \protect\citealp{Pratt2005}, (e) \protect\citealp{Gioia2004}, (f)
  \protect\citealp{Donahue2003}, (g) \protect\citealp{Tozzi2003} (h)
  \protect\citealp{Ellis2002}, (i) \protect\citealp{Worrall2003}, (j)
  \protect\citealp{Dressler1992}.}
\label{tab:coord}
\end{table*}

Our four chosen clusters have been studied at radio wavelengths by a
variety of groups employing wide--ranging techniques.  In the
subsequent sections we summarise previous work in terms of the most
recent, any that is close to our observing frequency, and any that
employs a similar observing strategy. These clusters have also been
studied in detail in X--rays.  We note briefly the interesting
characteristics reported in the literature, paying particular
attention to those which may affect the SZ decrement.  We also report
detections of lensing by our target clusters where appropriate.

\subsection{\cl}

\cl\space was observed by the pioneers of SZ astronomy, and continues
to be a popular target.  The SuZIE team performed simultaneous
bolometric observations in three frequency bands (145, 221 and
355\,GHz) and find a central Comptonisation parameter $y_0$ of
$(3.27^{+1.45}_{-2.86}) \times 10^{-4}$ \citep{Benson2004}, thus
$\Delta T_0 = -1785 ^{+792}_{-1616}\,\mu$K.  The work of
\cite{Tsuboi2004} is interesting, as they employ similar observing
techniques to those described in this work.  They find a central
decrement of $-1431\pm{133}\,\mu$K using the Nobeyama 45-m telescope,
at 43\,GHz.  \cite{Reese2002} find a similar result of
$-1242\pm{105}\,\mu$K, observing with the OVRO and BIMA
interferometers at 30\,GHz.

\cite{Worrall2003} present analysis of an \emph{XMM-Newton}
observation of \cl.  They find a best fitting temperature of
$9.13^{+0.24}_{-0.22}$\,keV within 1.5\,arcmin of the cluster centre
(see Table \ref{tab:coord}), and note the existence of a central
structure which may have a harder spectrum.  This, along with the
presence of an emission region to the west, may be evidence of merger
activity.  They comment that their determination of the total
gravitating mass is in good agreement with the value derived from
lensing data \citep{Smail1997}, thus supporting assumptions of
hydrostatic equilibrium.  Combination of their X-ray results with a
reanalysis of the SZ data presented in \cite{Hughes1998} yields a
value for the Hubble constant of $68\pm8\,\mathrm{km s^{-1}
Mpc^{-1}}$ (random error only).

\subsection{\four}

\cite{Benson2004} also observed \four, finding a central
Comptonisation parameter from their multi--frequency data of $(2.84\pm0.52)
\times 10^{-4}$, corresponding to $\Delta T_0 = -1550 \pm 284 \,
\mu$K.  This cluster was also included in the recent work of
\cite{Tsuboi2004}, who record a central decrement of $-1201\pm184\,\mu$K.
\four\space also features in the work of the OVRO/BIMA collaboration.
\cite{Reese2002} quote a central decrement from their 30\,GHz
measurements of $-1431^{+98}_{-93}\,\mu$K.

The gas temperature and luminosity of \four\space are given in Table
\ref{tab:coord}. \cite{Donahue2003} analyse a Chandra
observation in order to obtain a single temperature.  They extract
spectra from annular regions and do not find a significant gradient
between the inner and outer regions of the cluster using these data.
However, they also find an acceptable fit to a two--temperature plasma
model, although the associated cool component would not contribute
significantly to the X--ray emission.  \cite{Sand2005} identify
6 candidate lensing arcs in \four.

\subsection{\ten}

\ten\space is less well-studied at radio wavelengths.  It was observed
recently by \cite{Benson2004}, who find the central Comptonisation
parameter to be $y_0 = (3.87^{+1.19}_{-1.12}) \times 10^{-4}$ from the
measurements at 145\,GHz. This corresponds to a temperature decrement
of $2113^{+650}_{-612} \, \mu$K.  This cluster has also been observed
at 30\,GHz using the OVRO/BIMA arrays.  \cite{Joy2001} use the SZ data
to infer a cluster temperature of $10.4^{+5.0}_{-2.0}$\,keV, whereas
\cite{Grego2001} calculate the gas mass fraction within $r_{500}$
(where $r_{500}$ is the radius inside which the cluster density is
greater than 500 times the critical density) to be
$f_{\mathrm{g}}=0.053\pm0.028$.

\cite{Jee2005} present a joint analysis of Chandra and HST weak
lensing data for this cluster.  They find three dominant clumps in the
dark matter distribution, and observe that one substructure is not
present in the X-ray data.  The remaining two features are displaced
in the X-rays relative to the HST shear map, probably as a result of
a current merger.  They find consistent mass estimates ($r<1$\,Mpc),
using the datasets independently, of $\sim 1 \times
10^{15}\,M_{\odot}$.  They also report a significant temperature
gradient, but comment that no shock--heated region is observed between
the X-ray peaks.

\subsection{Abell 2218}
\label{sec:A2218}

A2218 was one of the first clusters to be detected via the SZ effect
in the 1970s, and has been studied in much detail since.  Most
recently, \cite{Jones2005} observed the cluster at 15\,GHz with the
Ryle telescope, and found a maximum central decrement of $-760\pm150
\, \mu$K.  \cite{Reese2002} record a central decrement of
$-731^{+125}_{-150} \, \mu$K.  \cite{Tsuboi1998} observed this
cluster, this time at 36\,GHz, and measured the SZ effect to be
$-680\pm{190} \, \mu$K.

A2218 has been studied extensively at X--ray wavelengths, and has an
X-ray luminosity of $9.55 \times 10^{44}\,$erg\,s$^{-2}$ and an
average gas temperature of $6.63\pm0.27$\,keV.  A Chandra study of BCS
clusters including A2218 is presented in \cite{Bauer2005}.  They find
that the cluster exhibits a cool core.
This cluster is also part of a HST lensing study
(\citealp{Smith2005}).  The authors combine strong and weak lensing
information in order to deduce a total cluster mass of
$(5.6\pm0.1)\times10^{14}M_{\odot}$.  Using additional information
from archival Chandra observations, they compare the mass and X-ray
morphology of the cluster to establish quantitatively that it appears
dynamically immature. This is supported by other Chandra studies, such
as that reported in \cite{Govoni2004}.  This work reinforces previous
claims that A2218 has an irregular temperature structure, despite its
relatively symmetric X-ray morphology.


\section{OBSERVATIONS AND DATA REDUCTION}

\subsection{Observing Strategy}
\label{sec:obs}

The observations presented in this paper were performed at intervals
between June 2005 and April 2006.  We aimed to achieve a similar noise
level for all clusters, thus observing times were roughly 11 hours for
each.  The time spent on the cluster itself was less than half of this
value due to telescope control issues which have since been resolved.
Changes in the atmosphere are problematic for single--dish
experiments, and we employ a differencing technique in order to remove
the resulting effects.  For each cluster we make two observations,
firstly the cluster itself (hereafter referred to as the TARGET field)
and secondly a trailing field (hereafter referred to as the TRAIL
field) at the same declination but offset in right ascension.  To this
TARGET--TRAIL strategy we apply both beam switching and position
switching.

Firstly for the TARGET field, the horn--feeds are positioned such that
one beam, beam $A$, is coincident with the cluster centre and the
other, beam $B$ provides a measure of the blank sky signal.  (For
extended sources, beam $B$ may measure a small signal itself.  This
must be properly accounted for - see Section \ref{sec:beta}.) We
switch between the $A$ and $B$ beams at a rate of 280\,kHz, recording
the integrated $A-B$ difference every second.  The beams are
repositioned every 53 seconds such that beam $B$ measures the cluster
and beam $A$ measures the sky background, and the differencing is
repeated.  We then take the double--difference, i.e. $(A_1-B_1) -
(B_2-A_2)$, such that we recover twice the cluster signal relative to
the two background regions.  This process is then repeated for a TRAIL
field, which is observed over the same range of hour angle as the
TARGET field.  We are satisfied that our TRAIL fields contain at worst
negligible amounts of contaminant signal and thus retain the data
purely as a check.

\subsection{Calibration}

The data are calibrated against an internal noise source, which is
itself calibrated via observations of the well--known bright radio
sources NGC7027 and 3C286 with assumed flux densities 5.64\,Jy
\footnote{We note that this value, extrapolated to 30\,GHz via the
  scale of \cite{Baars1977}, is 3.5\% lower than the measurements
  of \cite{Mason1999}.  This is well below our systematic error level of
  5\%. } and 2.51\,Jy. Significant changes in the level of the
  internal noise source occurred over the duration of these
  observations. Some were due to changes in the way that the noise
  source was used, but others were due to a temperature sensitivity of
  the noise source, which was controlled by improving the temperature
  stability in the receiver cabin. Residual uncertainties in the
  calibration of the system at the level of 5\% may be expected.

\subsection{Statistical Data Analysis}

After combination of the second-by-second average data into
double--differenced measurements of the brightness of the sky, and
calibration, the data are examined for periods of increased noise
(which might arise from receiver instabilities or bad weather
conditions) or individual anomalous points. The quality of the data
throughout the observing period was good, with a tendency to show a
slight increase in noise in Spring 2006 as the weather became warm and
wet: weather effects were minimized by the bulk of the observations
being taken in visually clear conditions.

The combination of the data into final averages was performed
including statistical tests for outlier data. The fractions of the data
points rejected by $3\sigma$ or $5\sigma$ cuts were small in all
cases, and no cut-dependent changes in the average results were
seen. The distributions of data values in the double-difference data
are close to Gaussian, with a slight tendency to show elevated wings
in the distributions: the estimates of the error on the mean
(Table \ref{results}) take account of these deviations from Gaussian
distributions.

\subsection{Removal of Instrumental Signature}
\label{sec:beta}

As mentioned in Section \ref{sec:obs}, the small separation of the
OCRA-p beams relative to the extent of a cluster's SZ decrement leads
to a significant reduction in the measured signal compared to the
intrinsic central signal.  In order to remove this effect and recover
the `true' SZ signal, we adopt the standard $\beta$--model approach.
We take values for the core radius $r_{\mathrm{c}}$ and $\beta$ from
the literature (see Table \ref{tab:beta}) in order to estimate the
fraction, $F$ of the SZ signal measured by beams $A$ and $B$
respectively, given where the beams fall on the sky, and thus the
fraction of the true signal which we expect to measure with the OCRA-p
system.  We can then estimate the SZ signal which would be measured by
an `ideal' instrument with narrow beams at infinite separation.  (We
note that the spherical model is non--ideal, but argue that in this
context the adverse effects are small given that our observations are
restricted to the central regions of the clusters.  For more detailed
analysis in the future we will adopt more realistic models, but do not
feel that it is necessary here).  We also account for the parallactic
angle range for each set of observations.  The results are recorded in
Table \ref{tab:beta}.  It is interesting to note that $F$ is very
similar (and close to 0.5) for all clusters in the sample despite the
large range of redshifts involved.  This flatness of response is a
generic feature of many practical SZ observing techniques (see,
e.g., \citealp{Birkinshaw2005}).

\begin{table*}
\begin{tabular}{lr@{}lr@{}lccc}
\hline 
Cluster &\multicolumn{2}{c}{$r_{\mathrm{c}}$} &\multicolumn{2}{c}{$\beta$} &Offset &$F$ &Refs\\
 &\multicolumn{2}{c}{(arcsec)} & & &(arcsec) & \\
\hline
\cl &$36.6${}&$\pm1.1$ &$0.697${}&$\pm0.010$ &30.9 &0.52 &a\\ 
\four &$45.1${}&$\pm1.2$ &$0.88${}&$\pm0.02$ &17.6 &0.53 &b\\
\ten &$53.4${}&$^{+22.2}_{-12.6}$ &$0.96${}&$^{+0.48}_{-0.22}$ &21.7 &0.55 &c\\
A2218 &$56.0${}&$\pm0.84$ &$0.591${}&$\pm0.004$ &28.4 &0.43 &b\\
\hline
\end{tabular}
\caption{$\beta$--model parameters, estimated offset of pointing
centre from apparent X-ray centre, and the estimated fraction of SZ
flux measured by the OCRA-p two--beam system. $\beta$--model
parameters taken from (a) \protect\cite{Worrall2003}, (b)
\protect\cite{DeFilippis2005}, (c) \protect\cite{Neumann2000}}
\label{tab:beta}
\end{table*}


\section{RADIO SOURCE CONTAMINATION}
\label{sec:src}

Radio source contamination remains a significant problem for Sunyaev
Zel'dovich observations.  Here interferometers have a distinct
advantage over single--dish experiments as they offer the possibility
of Fourier filtering.  Techniques for implementing this method are
discussed in \cite{Grainger2002} and \cite{Reese2002}.  For an
instrument such as OCRA-p which operates a position switching
strategy, radio sources can affect the data by producing a
\emph{positive} signal when they lie close to the cluster centre, or a
rogue \emph{negative} signal when they lie in the reference beam of
the telescope. If we know the source flux densities at our observing
frequency, we can correct the data and establish the true magnitude of
the SZ effect for each cluster.

No high--frequency survey of the radio sky exists, but for the four
well--known clusters used for our commissioning program we turn to the
extensive SZ work reported in the literature.  We are fortunate that
the OVRO/BIMA group have observed all clusters in our sample at
28.5\,GHz or 30\,GHz, or both.  We make use of the results reported in
\cite{Cooray1998}, \cite{Reese2002} and \cite{Coble2006}.  The primary
beams of OVRO and BIMA are $4'.2$ and $6'.6$ respectively, and so
cover the full fields of relevance for OCRA-p.  We also consult the
all--sky NVSS (\citealp{NVSS}, 1.4\,GHz) and GB6 (\citealp{GB6},
4.85\,GHz) catalogues, plus the 5\,GHz observations made by
\cite{Moffet1989}.  We consider all sources within 5' of the pointing
centre for each cluster.

\begin{figure*}
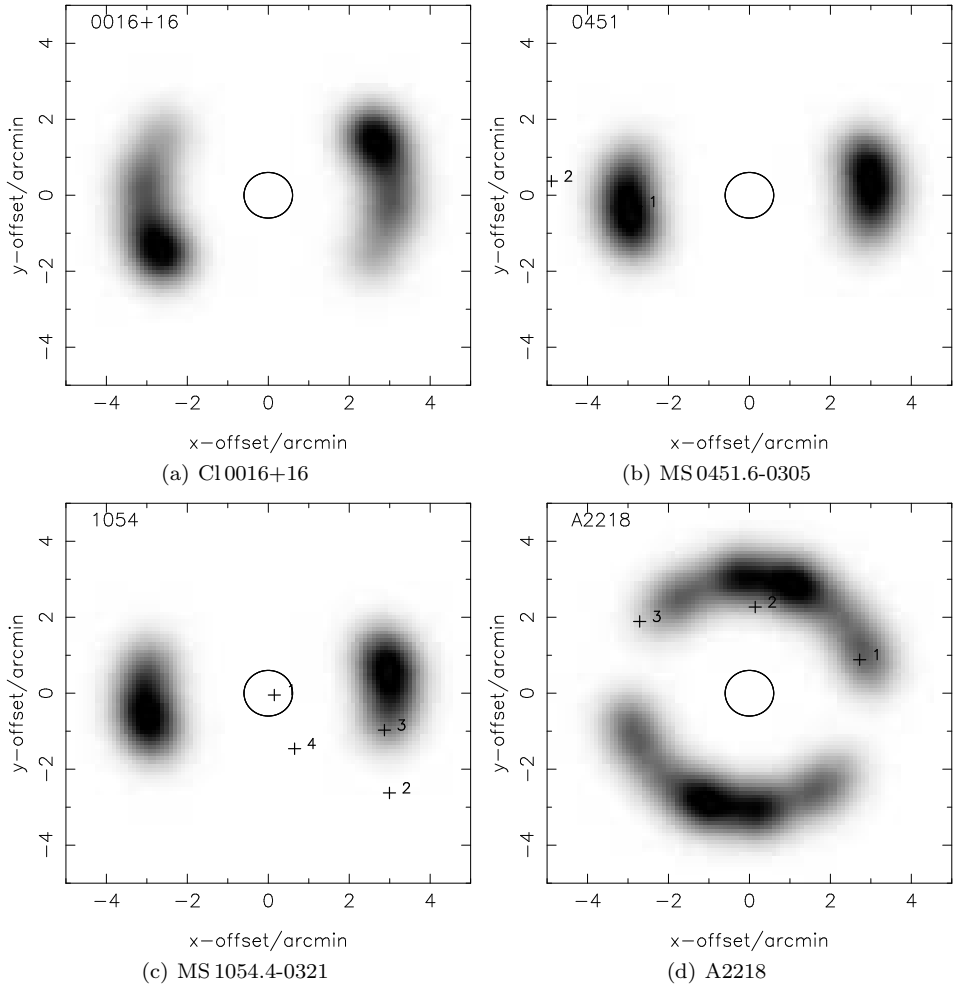

\subfigure[\cl]{
\includegraphics*[width=0.35\linewidth]{0016skyp.ps}}
\subfigure[\four]{
\includegraphics*[width=0.35\linewidth]{0451skyp.ps}}
\subfigure[\ten]{
\includegraphics*[width=0.35\linewidth]{MS1054skyp.ps}}
\subfigure[A2218]{
\includegraphics*[width=0.35\linewidth]{A2218skyp.ps}}
\caption{Radio sources in each cluster, shown relative to beams A and B,
  where the `off' beam B traces arcs as the telescope tracks the cluster.
  The source numbers correspond to the details presented in Table
  \ref{table:sources}.}
\label{figure:sources}
\end{figure*}

All potentially problematic sources are listed in Table
\ref{table:sources}.  Additionally, the source positions relative to
the telescope beams are plotted on RA--dec axes in Figure
\ref{figure:sources}.  The greyscale illustrates the arcs generated by
the moving blank--sky beams, and the density of points is proportional
to the observing time spent at each parallactic angle.  The central
circle marks the position of the `on' beam at all times.

We apply corrections to our SZ data based on the positions and flux
densities of the radio sources in each field.  Where we have a
measurement of a source at 30\,GHz, we use the value directly.  For
the sources for which information is available at two or more lower
frequencies, we fit a simple spectrum and extrapolate to estimate the
30\,GHz flux density.  Although this method is non--ideal, we note
that the error introduced by extrapolating from 28.5 to 30\,GHz is
likely to be very small.  Where only one measurement is available, we
adopt a cautious approach and run our correction software twice: once
using only the sources detected by OVRO/BIMA, and again using an upper
limit on the 30\,GHz flux.  Contamination through sources at 30\,GHz
is generally low for bright clusters, producing only small corrections
to the SZ data, and thus introducing little bias as a result of the
limitations of the methods employed.  We do not make any allowance for
source variability - this is discussed further in Section \ref{sec:res}.

We now examine the radio environment for each cluster in turn, and
outline the procedure adopted for each.

\subsection{\cl}

We find no sources within 5' of \cl\space and thus no correction to the SZ
measurement is required.

\subsection{\four}

For \four, source 1 lies inside the reference arc, and source 2
approximately one beamwidth outside.  We are fortunate to have a
30\,GHz flux density measurement for source 1 which we use directly.
Source 2 is detected only in the NVSS catalogue.  It is conceivable
that this source is lost in the noise of the OVRO/BIMA maps, and thus
we run our correction software twice: once without the additional
source, and again using an upper limit for its flux density.  The
value for the upper limit is taken as three times the RMS quoted for
source 1, $0.78\pm0.26$\,mJy.  Our measurements are unaffected by the
presence (or absence) of this source, since it lies outside the
reference arc and has a low flux density.

\subsection{\ten}

We find four possible contaminant sources in the \ten\space field.
Source 1 lies close to the pointing centre.  Source 3 is found inside
the reference arc so may also be a problem, although it will enter
with low weight.  Sources 2 and 4 lie outside and inside the reference
arcs respectively.  We have 28.5\,GHz measurements for sources
1, 3, and 4 which we combine with the NVSS values at 1.4\,GHz in order
to estimate 30\,GHz fluxes as before.  For the remaining source, we
apply a similar approach to that used for \four.  We place an upper
limit of $0.12\pm0.04$\,mJy and re-run the source correction
procedure.  Again we find that the presence/absence of this source
does not affect our measurement, despite its position in the reference
arc, presumably to the low value of the upper limit placed on its flux
density.

\subsection{A2218}

A2218 has been observed at 30\,GHz, and three sources are detected.
We use these flux densities directly.

\begin{table*}
\begin{minipage}{150mm}
\setlength{\extrarowheight}{3pt}
\begin{tabular}{lclllllr@{}lc}
\hline
Cluster  &Source  &RA &DEC &$S_{30.0}$ &$S_{28.5}$ &$S_{5}$
&\multicolumn{2}{l}{$S_{1.4}$} &References\\
&No. &(J2000) &(J2000) &(mJy) &(mJy) &(mJy) &\multicolumn{2}{l}{(mJy)} &\\
\hline
\cl &- &- &- &- &- &- &- &- &-\\
\hline
\four &1 &04 54 22.1 &-03 01 25  &$1.41{}\pm0.26$
   &$1.86{}\pm0.26$ &... &$14.9$&${}\pm0.7$ &a, a, -, c\\
 &2 &04 54 30.53 &-03 00 44.5 &... &... &... &$14.6$&${}\pm0.6$ &-, -, -, c\\
\hline
\ten  &1 &10 56 59.6 &-03 37 30 &... &$0.90{}\pm0.04$
&... &$14.1$&${}\pm0.9$ &-, d, -, c\\
 &2 &10 56 48.24 &-03 40 04.6
&... &... &... &$4.6$&${}\pm0.4$ &-, -, -, c\\
 &3 &10 56 48.74 &-03 37 25.4
&... &$1.4\pm{}0.04$ &... &$18.2$&${}\pm1.0$ &-, d, -, c\\
 &4 &10 56 57.58 &-03 38 54.7
&... &$0.4\pm{}0.04$ &... &$3.1$&${}\pm0.4$ &-, d, -, c\\
\hline
A2218  &1 &16 35 22.1 &66 13 23 &$4.29\pm{}0.21$
   &$4.43{}\pm0.20$ &$2.8{}\pm0.2$ &$1.25$&${}\pm0.21$ & a, a, b, b\\
       &2 &16 35 47.7 &66 14 46 &$1.36{}\pm0.10$
   &$1.59{}\pm0.11$ &$3.7{}\pm0.3$ &$18.0$&${}\pm1.8$ &a, a, b, c\\
 &3 &16 36 16.0 &66 14 23 &$2.41{}\pm0.29$ &$3.13{}\pm0.30$
  &$4.2{}\pm0.1$
&$13.3$&${}\pm0.6$ &a, a, b, c\\
\hline
\end{tabular}
\setlength{\extrarowheight}{0pt}
\caption{Radio sources associated with each of the six clusters.
  Fluxes taken from (a) \protect\cite{Reese2002}, (b)
  \protect\cite{Moffet1989}, (c) \protect\cite{NVSS}, (d)
  \protect\cite{Coble2006}, (e) 4.85\,GHz, \protect\cite{GB6}.
  The source index refers to Figure \ref{figure:sources}.}
\label{table:sources}
\end{minipage}
\end{table*}

\section{RESULTS AND DISCUSSION}
\label{sec:res}

\begin{figure*}
\subfigure[\cl]{
\includegraphics*[width=0.3\linewidth, angle=-90]{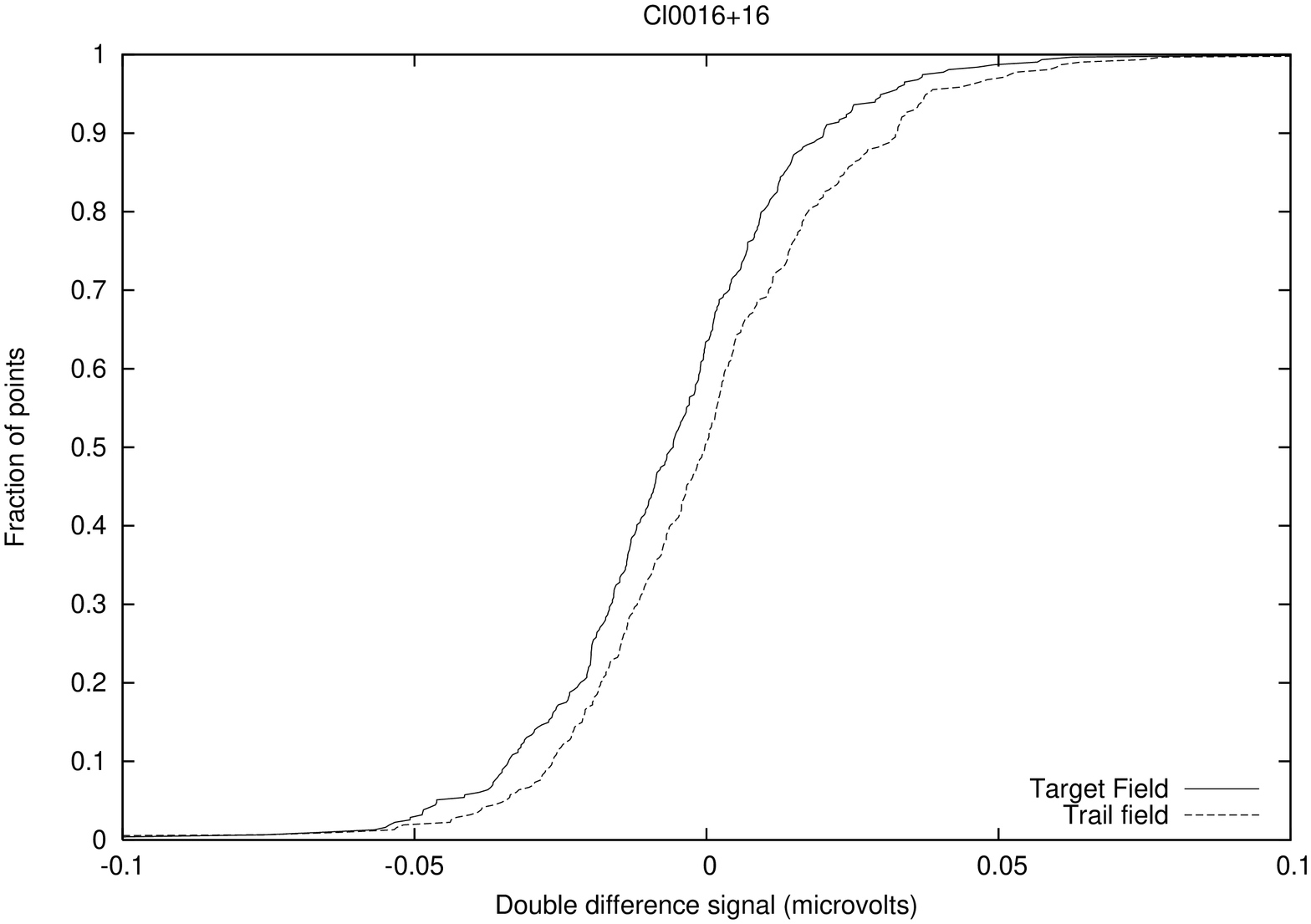}}
\subfigure[\four]{
\includegraphics*[width=0.3\linewidth, angle=-90]{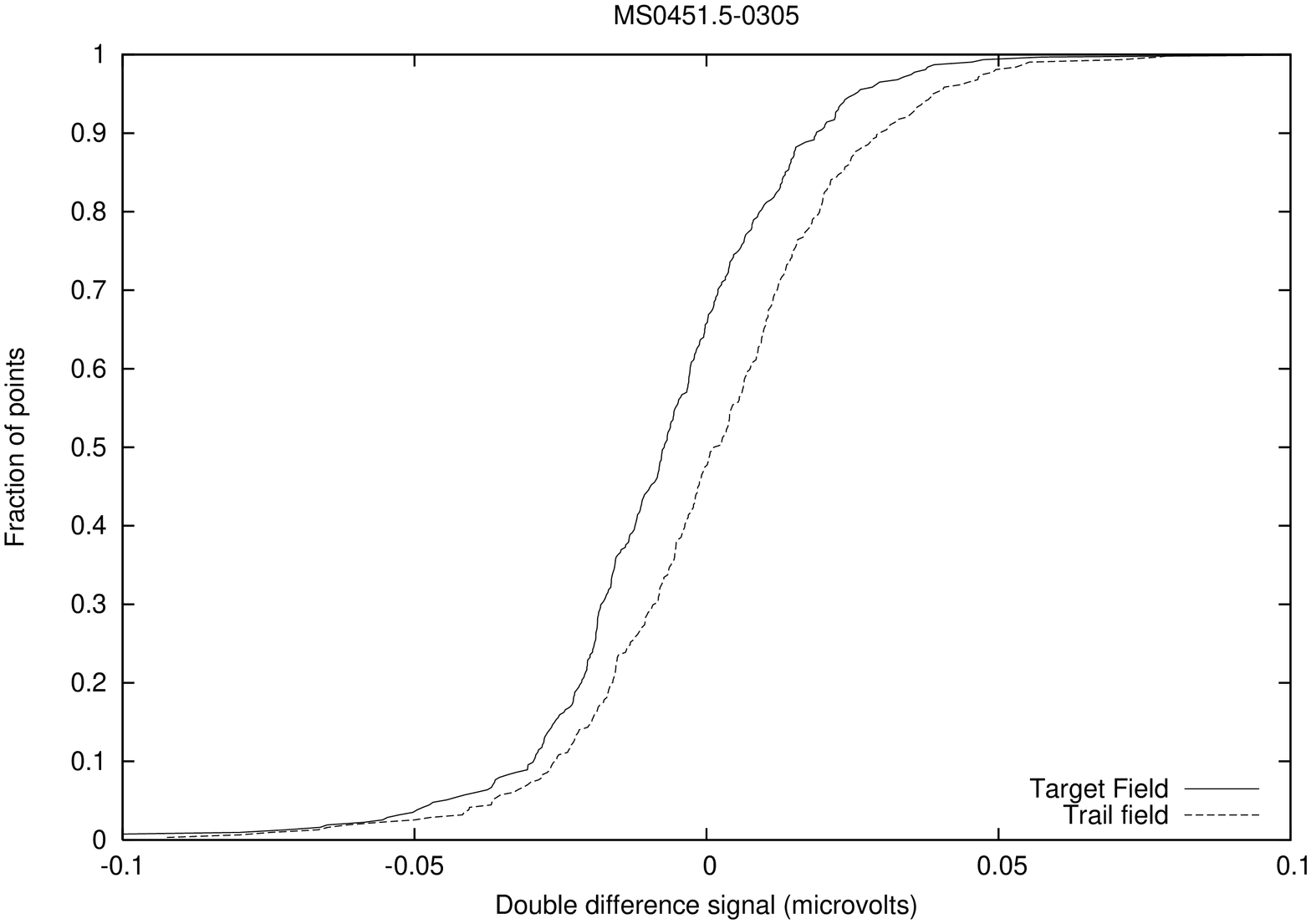}}
\subfigure[\ten]{
\includegraphics*[width=0.3\linewidth, angle=-90]{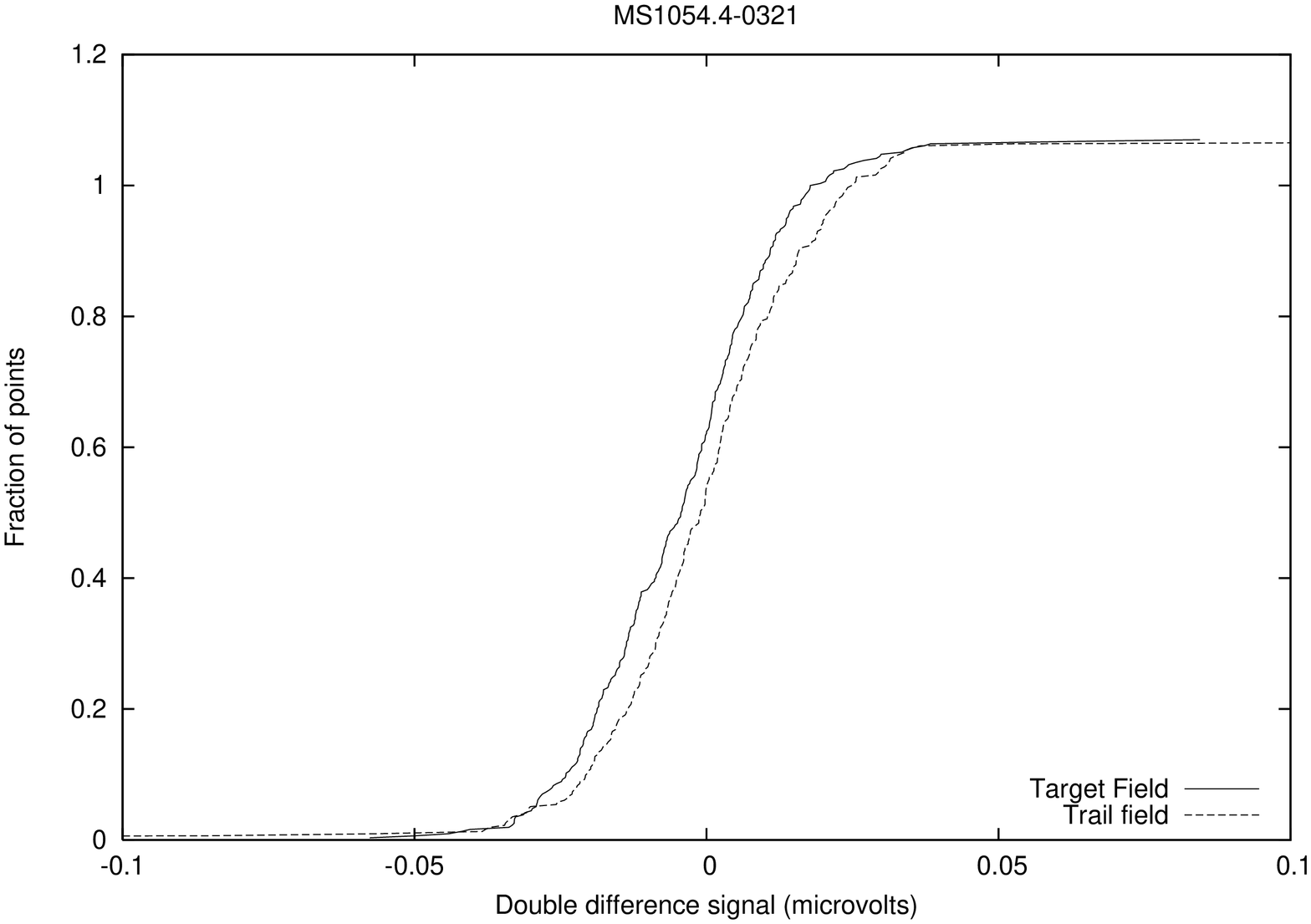}}
\subfigure[A2218]{
\includegraphics*[width=0.3\linewidth, angle=-90]{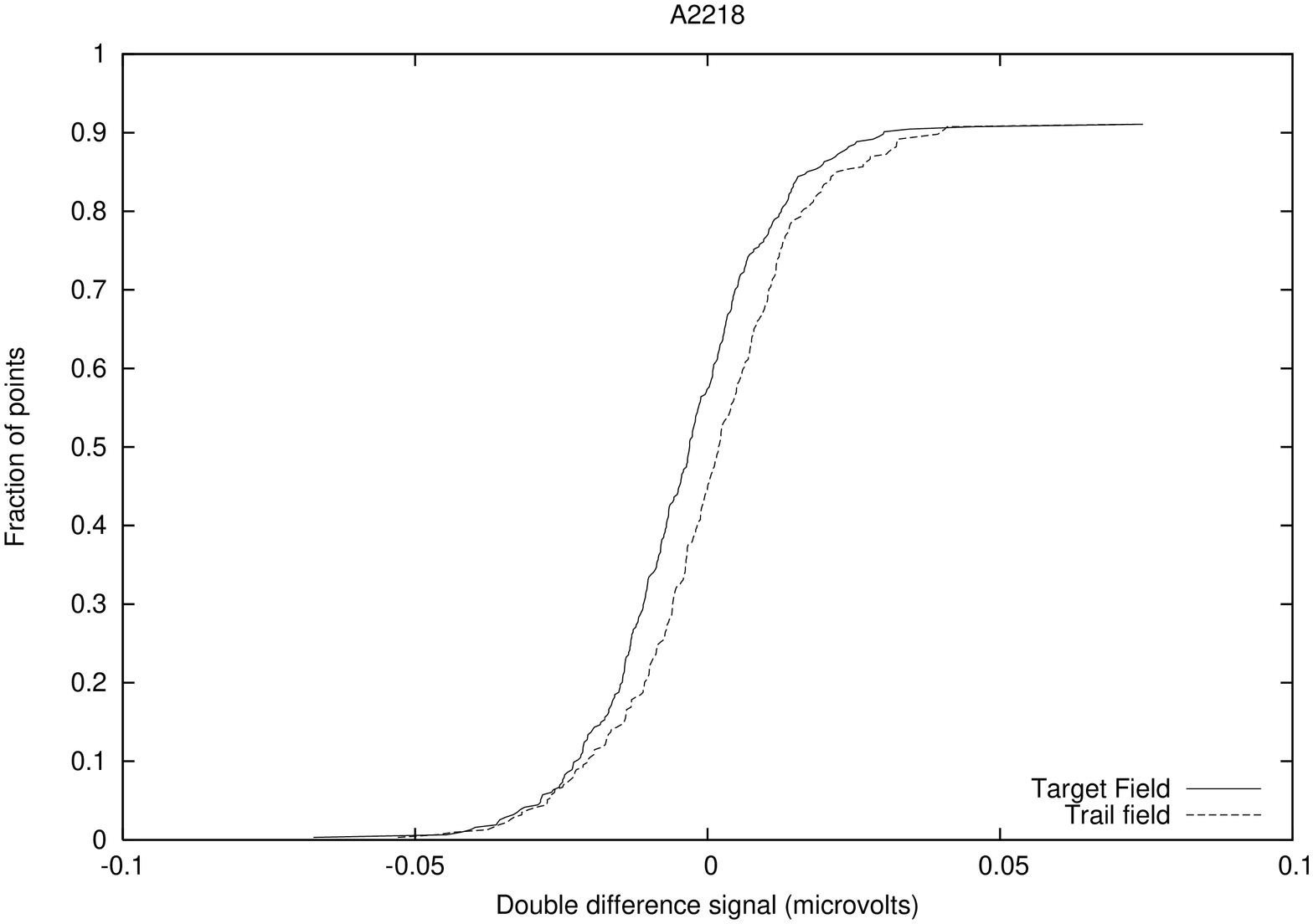}}
\caption{Cumulative histograms illustrating the raw data obtained through
  observing the Target and Trail fields for each cluster.  Note the
  negative shift of the Target curve relative to the Trail
  curve, indicating clear SZ detections.  In each case, the
  distributions are clearly separated by a standard
  Kolmogorov--Smirnov test.}
\label{figure:histograms}
\end{figure*}

Our SZ measurements are presented in Table \ref{results}.  Column two
contains the double--differenced values measured for the TARGET field
for each cluster.  Column three lists the same values for the TRAIL
fields, which we note are generally consistent with zero.  As
mentioned in Section \ref{sec:obs}, we feel that this eliminates the
need for our third level of differencing and subsequently will use the
TRAIL measurements as a check only.  We illustrate our data via the
cumulative histograms presented in Figure \ref{figure:histograms}.
For each cluster, the TARGET data are clearly shifted towards negative
values relative to the TRAIL data, indicating clear detections.  We
apply a standard two--sided Kolmogorov--Smirnov test and verify that
at the 1\% level, the measurements from the TARGET fields have
different distributions from the TRAIL fields.

The double--differenced values in column two of Table \ref{results}
represent only a fraction of the actual SZ signal (as discussed in
Section \ref{sec:obs}).  We list flux density values after correction
(see Section \ref{sec:beta} for the instrumental signature and radio
source contamination (see Section \ref{sec:src}) in column four.
Central temperature decrements are summarised in column five, with the
amount of telescope time listed in column six (note that this is
roughly twice the actual integration time).  We achieve clear
detections of our four well--known clusters.

\begin{table*}
\begin{tabular}{lcccccc}
\hline
Cluster &TARGET flux &TRAIL flux &$S_{\mathrm{SZ}}$ &$\Delta T$ &Telescope Time\\
 &(mJy) &(mJy) &(mJy) &($\mu$K) &(Hours)\\
\hline
\cl &$-1.884\pm0.345$ &$0.540\pm0.411$&$-4.28\pm0.78$ &$-1647\pm302$ &11.5\\
\four  &$-2.500\pm0.425$ &$0.446\pm0.347$ &$-4.05\pm0.80$ &$-1558\pm309$ &10.5\\
\ten &$-1.953\pm0.406$ &$0.331\pm0.384$  &$-4.48\pm0.74$ &$-1722\pm283$ &13.5\\
A2218 &$-1.696\pm0.317$ &$0.036\pm0.286$ &$-3.01\pm0.75$ &$-1159\pm288$ &11.0\\
\hline
\end{tabular}
\caption{Measured fluxes in the TARGET and TRAIL fields, difference
    and corrected measurements, and central temperature decrements.}
\label{results}
\end{table*}

To facilitate comparison, we also list our results alongside those
quoted in Section \ref{sec:clusters} in Table \ref{tab:comp}.  Our
results for \cl, \four, and \ten\space are in good agreement with
previous work, affirming that our observing strategy, radio source
treatment, and modeling are all effective.  The one exception is A2218
which is rather high, although only $\sim 1.5 \sigma$ away from the
other measurements.  However, as discussed in Section \ref{sec:A2218},
this cluster displays irregular morphology.  With the OCRA beam we may
be measuring the SZ signal from a particularly high pressure region.
Due to the extended nature of the SZ effect and ambiguities in
defining cluster centres, it is likely that our pointing centre
differs from those used by the comparison experiments.  Additionally,
we note that we have not incorporated the possibility of source
variability in our analysis.  We deduce that this is not a major
factor here, given the good agreement between our measurements and
those reported in the literature, although this could perhaps provide
an alternative explanation for the discrepancy noted above.

The short integration time required to obtain these clear detections
demonstrates the power of OCRA--p in SZ studies.  We stress that the
values quoted in table 4 correspond to the total time required to take
the data.  The OCRA control system is rather inefficient (although
recently improved) and thus the telescope spent more than half of this
time off source.  The predicted noise on a one second integration of
$\sim3.6$\,mJy degrades by a factor of $\sim2$ to the measured value
of $7.0$\,mJy.  Integrating over the full `on--source' time of around
3 hours (8478 samples), we would expect this value to reduce to
$\sim0.08$\,mJy, which is better than the measured $\sim0.4$\,mJy by a
factor of 5.  This degradation is likely due to non--optimal removal
of atmospheric effects, which may partly be attributed to the poor
telescope control.  With improvements to both the OCRA receiver and
the telescope control system, we expect OCRA's sensitivity to increase
substantially in forthcoming observing runs.  Over the current season,
we have scheduled observations of a large, statistically complete,
X--ray selected sample of clusters at $z>0.2$.  These observations
will form an SZ catalogue with uniquely well--understood selection
effects. This will enable us to test SZ/X--ray scaling relations
(e.g. \citealp{Cooray1999}).  The SZ measurements will be combined
with high--quality X-ray data in order to further test cluster models.
We also anticipate performing observations of high--redshift cluster
candidates found by the {\it XMM-Newton} Large Scale Structure survey
such as that reported by \cite{Bremer2006}.  The two samples will be
at significantly different mean redshifts, thus facilitating studies
of the evolution of cluster properties.  The OCRA--p commissioning
observations also pave the way for SZ imaging.  It is possible, by
means of numerous pointings, to map clusters using OCRA--p.  However,
a multiple--beam instrument such as OCRA--F would be far more
efficient.  The installation of OCRA--F on the telescope will allow us
to make the first well--resolved maps of SZ clusters with OCRA--F
after summer 2007.

OCRA--F, and indeed the full, 100--beam instrument, will also be used
to perform blind surveys for galaxy clusters. Some source correction
will be possible via follow--up observations of suspected sources
present in the reference beam traces.  Deeper observations, coupled
with reference arc analysis, would. allow contaminant sources to be
detected and removed provided that the data are taken over a
sufficiently large range of parallactic angles.  Naturally, there may
be some cases where sufficient additional information about possible
contaminant sources is not available. For clusters as massive as the
four discussed here, we do not expect radio sources to preclude
detection. If we were to ignore the source corrections, all four
clusters would still be unambiguously detected, and the temperature
decrements would remain close to the values reported in the
literature.  \cl\space has no source contamination so is unchanged.
The $\Delta T$ values for \four\space and A2218 become rather high as a
result of switching onto sources in the reference arcs ($-1814\pm308\,
\mu$K and $-1517\pm284\, \mu$K respectively) while \ten\space is low
due to the central radio source ($-1366\pm284\,\mu$K).  

As we move towards observing fainter clusters, we recognise that
contamination by primordial CMB features could potentially be a
problem, as noted in other work (e.g. \citealp{Lancaster2005}).
However, OCRA is a relatively high--resolution instrument, probing
arcminute scales which correspond to $3500\lesssim \ell \lesssim
10000$.   At such high multipoles, for a concordance cosmology and
including the effects of lensing, we expect the amplitude of the
temperature power spectrum to be of the order of a few $\mu$K.  This
would be well below the OCRA noise in surveys of scale $\sim100$
square degrees with planned sensitivity of $\sim100 \, \mu$K.

\begin{table*}
\begin{tabular}{ccccl}
\hline
\multicolumn{5}{c}{$\Delta T\,(\mu K)$}\\
\cl &\four &\ten &A2218 &References \\
\hline
$-1647^{+302}_{-302}$ &$-1558^{+309}_{-309}$ &$-1722^{+283}_{-283}$ &$-1159^{+288}_{-288}$  &OCRA \\
$-1785^{+792}_{-1616}$& $-1550^{+284}_{-284}$ &$-2113^{+650}_{-612}$ &- &\cite{Benson2004}  \\
$-1431^{+133}_{-133}$ &$-1201^{+184}_{-184}$ &- &$-680^{+190}_{-190}$ &\cite{Tsuboi2004}  \\
$-1242^{+105}_{-105}$ &$-1431^{+98}_{-93}$ &- &$-731^{+125}_{-150}$ &\cite{Reese2002}  \\
-&-&-&$-760^{+150}_{-150}$ &\cite{Jones2005} \\
\hline
\end{tabular}
\caption{Comparison of OCRA results for the four clusters with
temperature decrements quoted in the literature.}
\label{tab:comp}
\end{table*}

\section{ACKNOWLEDGMENTS}

We acknowledge support for the design and construction of OCRA-p from
the Royal Society Paul Instrument Fund, and funds for the data
acquisition system and operation on the telescope from the Ministry of
Science in Poland via grant number KBN 5 P03D 024 21 who, along with
PPARC, also supported the scientific exploitation of the completed
system.  We thank Anze Slosar for useful discussions. Finally, we
thank the anonymous referee for their valuable input.


\bibliographystyle{mn2e.bst}
\bibliography{ref_lancaster}


\bsp 

\label{lastpage}

\end{document}